\documentclass[10pt,twocolumn,letterpaper]{article}

\usepackage{iccv}
\usepackage{times}
\usepackage{epsfig}
\usepackage{graphicx}
\usepackage{amsmath}
\usepackage{amssymb}

\usepackage[pagebackref=true,breaklinks=true,letterpaper=true,colorlinks,bookmarks=false]{hyperref}

\iccvfinalcopy 

\ificcvfinal\fi

\begin{document}

\title{Uncertainty-aware GAN with Adaptive Loss for Robust MRI Image Enhancement}

\author{Uddeshya Upadhyay\\
IIT-Bombay\\
{\tt\small uddeshya@cse.iitb.ac.in}
\and
Viswanath P. Sudarshan\\
TCS Research\\
{\tt\tiny viswanath.pamulakantysudarshan@tcs.com}
\and
Suyash P. Awate\\
IIT-Bombay\\
{\tt\small suyash@cse.iitb.ac.in}
}

\maketitle
\ificcvfinal\thispagestyle{empty}\fi

\begin{abstract}
    Image-to-image translation is an ill-posed problem as unique one-to-one mapping may not exist between the source and target images. Learning-based methods proposed in this context often evaluate the performance on test data that is similar to the training data, which may be impractical. 
    This demands robust methods that can quantify uncertainty in the prediction for making informed decisions, especially for critical areas such as medical imaging. Recent works that employ conditional generative adversarial networks (GANs) have shown improved performance in learning photo-realistic image-to-image mappings between the source and the target images. However, these methods do not focus on (i)~robustness of the models to out-of-distribution (OOD)-noisy data and (ii)~uncertainty quantification. This paper proposes a GAN-based framework that
    (i)~models an adaptive loss function for robustness to OOD-noisy data that automatically tunes the spatially varying norm for penalizing the residuals and
    (ii)~estimates the per-voxel uncertainty in the predictions. We demonstrate our method on two key applications in medical imaging:
    (i)~undersampled magnetic resonance imaging (MRI) reconstruction
    (ii)~MRI modality propagation. 
    Our experiments with two different real-world datasets show that the proposed method (i)~is robust to OOD-noisy test data and provides improved accuracy and (ii)~quantifies voxel-level uncertainty in the predictions.
 \end{abstract}
 
 \section{Introduction and Related Work}
 The image-to-image translation methods that generate images of the target view, given images of the source view, are often \textit{ill-posed} problems as unique one-to-one mapping may not exist between the source and target views. Learning-based methods proposed in this context~\cite{ledig2017photo,liu2017unsupervised,huang2018multimodal} evaluate the performance on test data that is similar to training data. 
 Such models suffer from severe degradation in the performance when presented with out-of-distribution (OOD)-noisy data~\cite{Hsu_2020_CVPR,lee2018simple,moosavi2017universal}.
 In critical applications, like in medical imaging, in addition to robustness to OOD-noisy data, it is important to quantify the uncertainty in the predictions made by the model to aid clinical decisions~\cite{uncer_talarbel,sjolund2018bayesian}. 
 Particularly, problems such as enhancing the quality of a given medical image and synthesizing medical images of target modality given a source modality benefit from the uncertainty estimation by allowing risk assessment in the predicted images~\cite{UncerdMRI,uncer_talarbel,sjolund2018bayesian}. 
 In this work, we use two critical tasks in medical image analysis that can broadly be posed as image-to-image translation problems to demonstrate the efficacy of the proposed methods: 
 (i)~undersampled magnetic resonance imaging (MRI) reconstruction and
 (ii)~MRI modality propagation, i.e., synthesizing T2 weighted (T2w) MRI from T1 weighted (T1w) MRI.
 
 MRI k-space (a 2D complex-valued space based on 2D Fourier transform of the slices) data acquisition is a time-consuming process as the speed is dependent on hardware and physiological constraints~\cite{moratal2008k,schlemper2017deep,fastMRI}. 
 Typically, faster acquisitions are realized by undersampling the k-space. However, it leads to blurring or aliasing effects in the MRI image. We formulate the reconstruction of high-quality MRI scans using undersampled k-space data as a translation from low-quality MRI (lqMR) to high-quality MRI (hqMR) scans
 (lqMR $\rightarrow$ hqMR).
 Similarly,
 modality propagation is of interest because routine clinical protocols acquire images with multiple contrasts which are critical for better diagnostics. Acquiring multiple contrasts increases scanning time. T1w MRI and T2w MRI are among the most commonly acquired contrasts and synthesis of T2w from T1w is posed as a translation problem (T1w $\rightarrow$ T2w).
 
 Recently, conditional generative adversarial networks (GANs) have shown substantially improved performance for the above-mentioned tasks in comparison to other learning-based methods~\cite{mia_book,nie2018medical,dar2019image,pet_ood,icml21_workshop,isbi19_srgan,miccai19_qegan}. Such methods condition the generator (that generates the target view) with a source view and the task of the discriminator is to discriminate between the generated target samples from the ground-truth target samples through adversarial loss term. Additionally, some methods also impose a cycle-consistency penalty to constraint the ill-posed translation problem~\cite{CycleGAN,liu2017unsupervised,huang2018multimodal}. 
 %
 Along with improved accuracy, 
 quantifying uncertainty can potentially aid in clinical decision making, especially in the presence of OOD-noisy data, while improving the accuracy of subsequent tasks such as image segmentation, 
 classification, etc~\cite{UncerdMRI}.
 %
 In the general context of medical imaging, OOD-noisy perturbations
 could be systemic (scanner-related) and/or physiological. 
 Hence, beyond reliable image synthesis during inference,
 a thorough analysis of the robustness of the network along with
 quantification of uncertainty in the predicted images is crucial for 
 clinical translation of synthesis frameworks \cite{wang2019aleatoric,nair2020exploring,ravi2019adversarial}.
 Designing learning strategies that are robust to large errors (i.e., outliers whose residuals follow heavy-tailed distributions) is a well-studied problem in statistics, optimization, and parameter estimation \cite{mb_ijcv,statLwS,robustStat,cox1979theoretical}. 
  
 Quasi norm-based loss function, $\ell_q$, use a generalized notion of the norm, where $q$ (often called the shape parameter) can be tuned empirically for robust learning \cite{sun2010secrets,adaptiveloss_cvpr2019}. Moreover, such empirical methods often require the $q$ to be fixed spatially.
 This work proposes a novel loss function that automatically learns the shape parameter $q$ that can vary spatially for the image dataset along with the scale parameter that allows us to estimate per-voxel uncertainty.
 
 \section{Methods}
 \label{sec:methods}
 \begin{figure}
    \centering
    \includegraphics[width=0.45\textwidth]{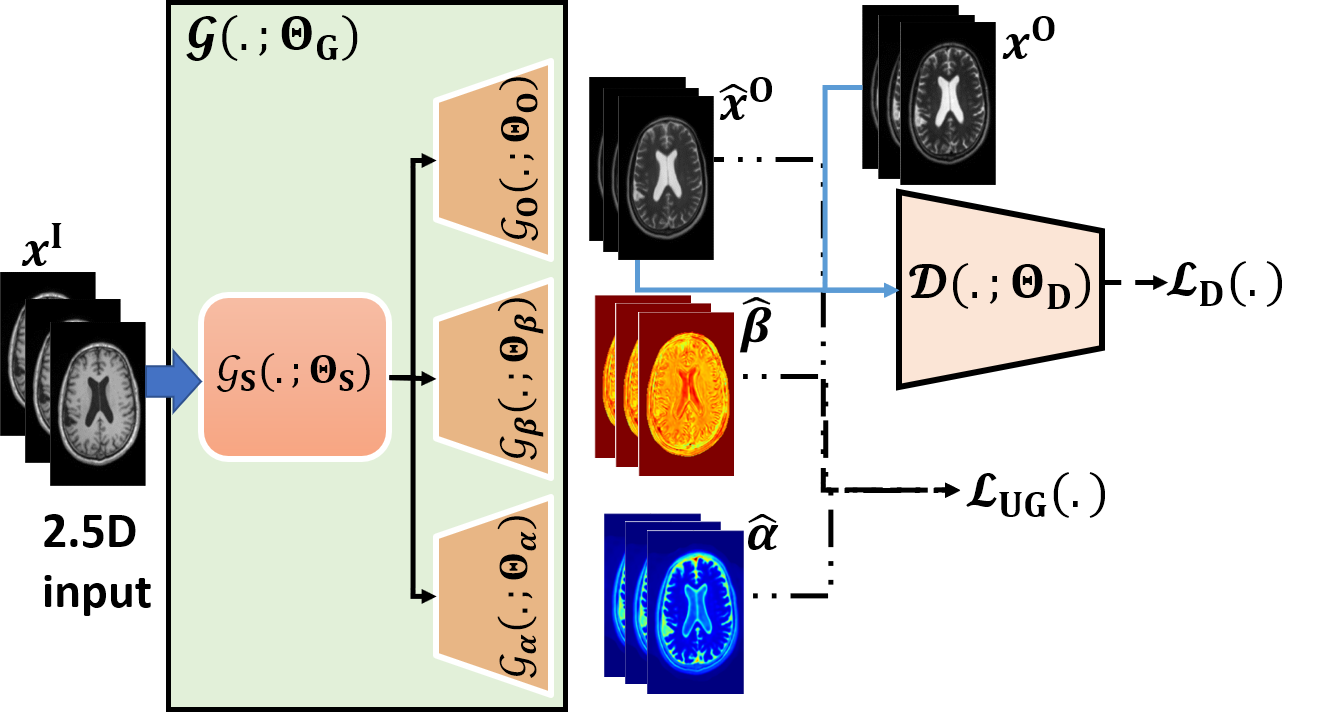}
    \caption
    {
    {\bf Model schema.} $\hat{x}^{\text{O}}$, $\hat{\beta}$, and $\hat{\alpha}$ denote the estimated output image, shape, and the scale parameter of the GGD prior respectively. 
    The parameters governing output $\hat{x}^{\text{O}}$ are a combination of the shared weights (across all three branches) in $\mathcal{G}_{\text{S}}$ and $\hat{x}^{\text{O}}$-specific weights in $\mathcal{G}_{\text{O}}$. Similarly,  $\hat{\alpha}$ and $\hat{\beta}$ depend on parameters in $\mathcal{G}_{\alpha}$ and $\mathcal{G}_{\beta}$, respectively, in addition to shared weights in $\mathcal{G}_{\text{O}}$ (refer Equation~\ref{eq:in1} -- \ref{eq:in3}).
    }
    \label{fig:model}
    \vspace{-15pt}
 \end{figure}
 We describe our proposed GAN framework and
 discuss specific details pertaining to the two applications: (i)~MRI reconstruction using undersampled k-space data posed as a quality enhancement (QE) task and (ii)~Modiality Propagation (MP). 
 Our proposed solution for these two applications can easily be extended to other image-to-image translation problems both with MRI as well as other modalities.
 
 Let the set of input images be represented by $X^{\text{I}}$ and the corresponding set of output images be represented by $X^{\text{O}}$.
 Let $x^{\text{I}}$ and $x^{\text{O}}$ represent elements from the set of input and output images (i.e., $x^{\text{I}} \in X^{\text{I}}$ and $x^{\text{O}} \in X^{\text{O}}$), respectively.  
 We learn the mapping from input to output using $T$ pairs of co-registered training data: 
 $\mathcal{X} := \{(x^{\text{I}}_t, x^{\text{O}}_t )\}_{t=1}^T$. 
 The input and output images for QE and MP are described below. 
 
 \textbf{Undersampled MRI reconstruction (QE).} Here, a mapping is learned from a low-quality MRI image (lqMR) (obtained via zero-filled inverse Fourier transform (IFT) of the undersampled k-space data) to the corresponding high-quality MRI image (hqMR) (obtained from fully-sampled k-space). 
 Therefore, for QE, $(X^{\text{I}}, X^{\text{O}})$ becomes $(X^{\text{LQ}}, X^{\text{HQ}})$.
 Let operator $\mathcal{F}$ represent the
 2D Fourier transform and matrix $H$ represent the
 sampling mask to perform undersampling in k-space. Then, the undersampled MRI is related to fully-sampled MRI via
 $
    x^{\text{LQ}} = \mathcal{F}^{-1}(\mathcal{F}(x^{\text{HQ}}) \odot H)
 $.
 With a fixed sampling mask, accounting for noisy k-space acquisitions, corrupted by complex Gaussian noise (say $\eta_{\text{K}}$), that represent OOD-noisy data~\cite{empEffectGauss_lustwig,zhu2020removal,circus}, the OOD-noisy input sample (say $x^{\text{LQ}}_{\text{OOD}}$) can be modeled by,
 \begin{equation}
    x^{\text{LQ}}_{\text{OOD}} = \mathcal{F}^{-1}((\mathcal{F}(x^{\text{HQ}}) + \eta_{\text{K}}) \odot H )
    \label{eq:k_corrupt}
 \end{equation}
 
 \textbf{Modality propagation (MP).} Here, we learn a mapping from T1w images to corresponding 
 T2w images. In this context, $(X^{\text{I}}, X^{\text{O}})$ can be represented 
 by the set of T1w MR and T2w MR images, i.e., $(X^{\text{T1}}, X^{\text{T2}})$.
 Unlike QE, the OOD-noisy input sample (say $x^{\text{T1}}_{\text{OOD}}$) during inference is obtained by adding Gaussian noise in the image-space (say $\eta_{\text{I}}$) to the T1w MRI input image (say $x^{\text{T1}}$).
 The addition of Gaussian noise in image-space is consistent with~\cite{mri_noise}, i.e.,
 \begin{equation}
    x^{\text{T1}}_{\text{OOD}} = x^{\text{T1}} + \eta_{\text{I}} 
    \label{eq:mri_corrupt}
 \end{equation}
 
 \subsection{Model}
 \label{ssec:models}
 %
 This work, inspired by conditional GANs, proposes a novel GAN-based framework which
 (i)~employs learnable quasi-norm loss, and 
 (ii)~estimates uncertainty maps, as described below.
 Let $\mathcal{G}(\cdot; \theta^{\text{G}})$ and $\mathcal{D}(\cdot; \theta^{\text{D}})$ represent 
 our generator and discriminator respectively.  
 The input to the generator is $x^\text{I} \in X^{\text{I}}$, 
 the predicted image is $\hat{x}^{\text{O}} = \mathcal{G}(x^{\text{I}}; \theta^{\text{G}})_O = \mathcal{G}_{\text{O}}(\mathcal{G}_{\text{S}}(x^{\text{I}}; \theta^\text{S});\theta^{\text{O}})$ 
 (i.e., output of $\text{O}$-branch of $\mathcal{G}$ as shown in Figure~\ref{fig:model}), and 
 the ground-truth is $x^\text{O} \in X^{\text{O}}$. 
 Each image consists of $K$ pixels. 
 We denote $j^{th}$ pixel in $i^{th}$ image, say $z$, as $z_{ij}$. 
 Prior works typically formulate the above task as a regression and the distribution of residuals 
 between the predictions and the ground-truths is assumed to be isotropic standard Gaussian distribution (zero mean and fixed standard deviation).
 However, some of the limitations of this approach include: 
 (i)~not being able to model the outliers/corruptions in the dataset, 
 as residuals due to outliers tend to follow heavy-tailed distributions.
 (ii)~the assumption of fixed standard deviation
 does not account for heteroscedastic uncertainty in their modeling. 
 Recent works model the heteroscedastic uncertainty by assuming that the variance is data-dependent~\cite{kendall}.
 We improve upon the assumptions made by prior works by
 (i)~modeling the residuals to follow a \textit{generalized Gaussian distribution} (GGD) with zero mean, which is 
 $
 \frac
 {
    \beta
 }
 {
    2\alpha \Gamma(\frac{1}{\beta})
 } 
 e^
 {
  -\left(\frac
       {
           |\epsilon-0|
       }
       {
           \alpha
       }
   \right)
   ^
   {
       \beta
   }
 }
 $,
 and (ii)~allowing the shape parameter ($\beta$) and the scale parameter ($\alpha$) to vary spatially.
 This allows us to learn appropriate quasi-norms at all spatial locations, 
 as well as estimating uncertainty at every output pixel. Therefore, 
 for the $i^{th}$ image, the residual at pixel location $j$,
 $\epsilon_{ij}$ (between the predicted value $\hat{x}^{\text{O}}_{ij}$ and ground-truth value $x^{\text{O}}_{ij}$), follows GGD, i.e.,
 \begin{align}
    \hat{x}^{\text{O}}_{ij} &:= x^{\text{O}}_{ij} + \epsilon_{ij} \\
    \epsilon_{ij} &\sim \frac{\beta_{ij}}{2\alpha_{ij} \Gamma(\frac{1}{\beta_{ij}})} e^{-\left(\frac{|\epsilon_{ij}-0|}{\alpha_{ij}}\right)^{\beta_{ij}}} \\
    \hat{x}^{\text{O}}_{ij} &\sim \frac{\beta_{ij}}{2\alpha_{ij} \Gamma(\frac{1}{\beta_{ij}})}e^{
        -\left(
            \frac{
                | ( \mathcal{G}_{\text{O}}(\mathcal{G}_{\text{S}}(x^{\text{I}}; \theta^\text{S});\theta^{\text{O}}))_{ij} - x^{\text{O}}_{ij}|
            }
            {
                \alpha_{ij}
            }
        \right)
        ^{ \beta_{ij} }
    }
 \end{align}
 \begin{figure}
    \centering
    \includegraphics[width=0.4\textwidth]{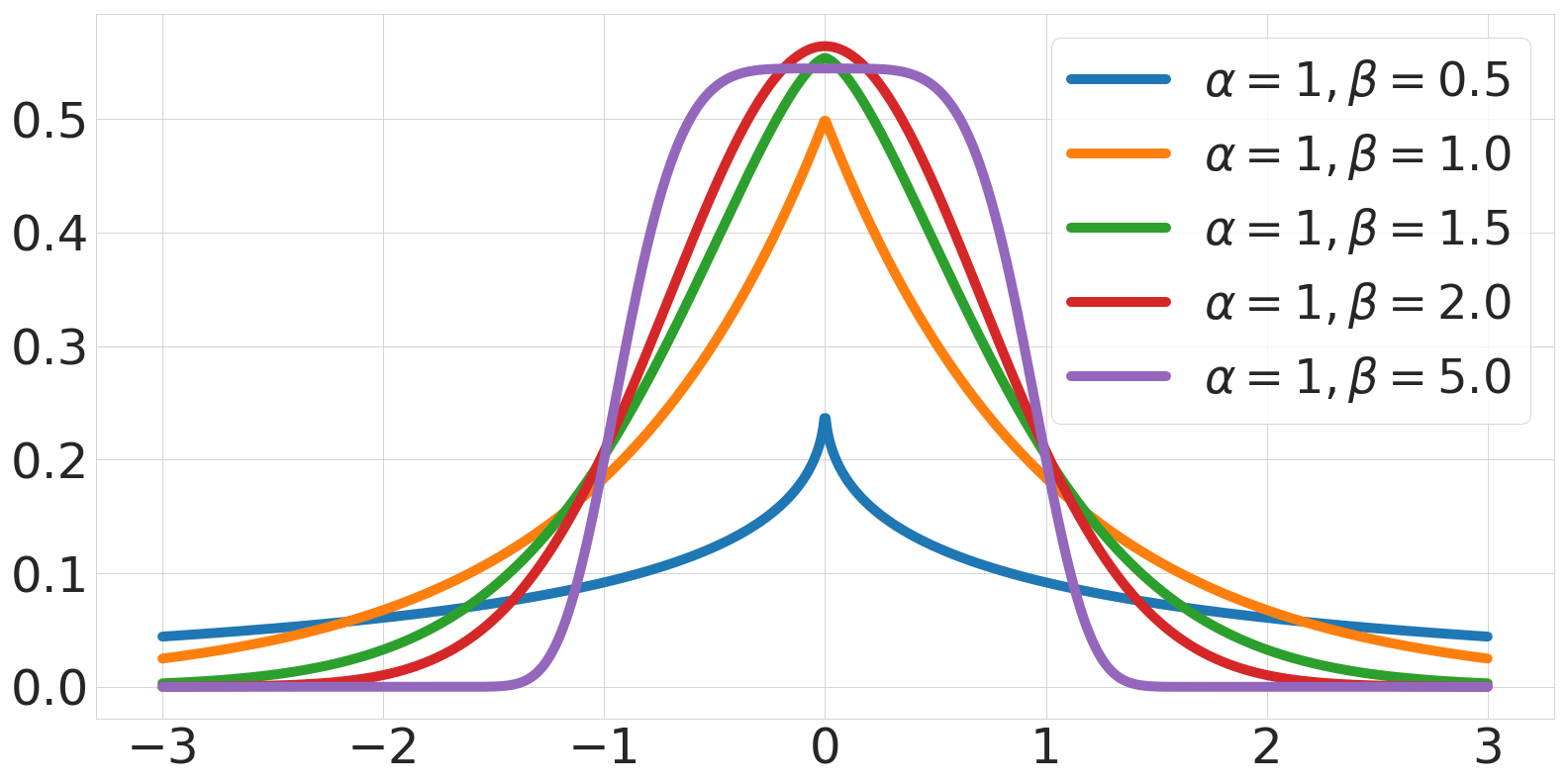}
    \caption
    {
    {\bf Generalized Gaussian Distribution}. PDF for GGD with differnt shape parameters ($\beta$). Lower $\beta$ corresponds to heavy-tailed distribution. 
    }
    \label{fig:ggd}
    \vspace{-10pt}
 \end{figure}
 GGD is capable of modelling heavy-tailed distributions including the Gaussian and Laplace PDFs as shown in Figure~\ref{fig:ggd}.
 Here $\alpha_{ij} > 0$ is the scale parameter, $\beta_{ij} > 0$ denotes the shape parameter, and $\Gamma(\cdot)$ is the standard gamma function.
 In our formulation all the $\epsilon_{ij}$'s are independent but not necessarily identically distributed as $\alpha_{ij}$ and $\beta_{ij}$ may vary spatially. 
 Hence, the likelihood is,
 \begin{align}
     & P(\mathcal{X}|\Theta) := 
     \prod_{i=1, j=1}^{i=T, j=K}
    \frac
     {\beta_{ij}}
     {2\alpha_{ij}\Gamma(\frac{1}{\beta_{ij}})}
     e^{
     \left (
         \frac{   
            -|
            ( \mathcal{G}(x^{\text{I}}; \theta^{\text{G}})_O)_{ij}
            -
            x^{\text{O}}_{ij}
            |
        }
        {
            \alpha_{ij}
        }
    \right )
    ^{ \beta_{ij} }
    } 
 \end{align}
 Therefore, log-likelihood is,
 \begin{align}
   & \log P(\mathcal{X}|\Theta) = \nonumber\\ 
   & \sum_{i=1, j=1}^{i=T, j=K} 
   -\left( 
       \frac{
           |\hat{x}^{\text{O}}_{ij} - x^{\text{O}}_{ij}|
       }
       {
           \alpha_{ij}
       } 
   \right)
   ^
   { \beta_{ij} }
   + 
   \log \frac{ \beta_{ij} } { 2\alpha_{ij} } 
   -
   \log \Gamma( \frac{1}{ \beta_{ij} } )
 \end{align}
 
 \begin{figure}
    \centering
    \includegraphics[width=0.4\textwidth]{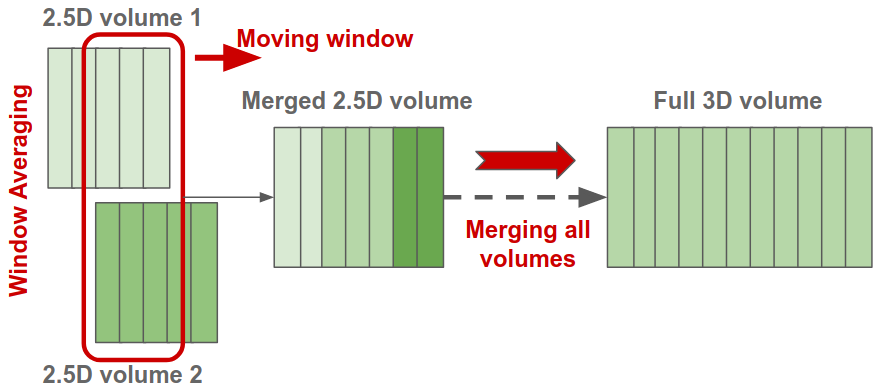}
    \caption
    {
    {\bf Merging scheme.} Moving window average to merge overlapping 2.5D volumes to produce a smooth 3D volume.
    }
    \label{fig:mergescheme}
    \vspace{-8pt}
 \end{figure}
 
 where $\Theta$ represents the collection of network parameters.
 In this way, to improve the robustness of the network, we predict 
 (i)~$\hat{x}^{\text{O}}_{ij}$ (the predicted value at every pixel location, also the mean of GGD), 
 (ii)~$\hat{\alpha}_{ij}$ (an estimate for the true $\alpha_{ij}$ at every pixel location), and 
 (iii)~$\hat{\beta}_{ij}$ (an estimate for the true $\beta_{ij}$ at every pixel location). Hence, the proposed robust quasi-norm based loss is
 \begin{align}
    \mathcal{L}_{\text{U}}(
        \{\hat{x}_{ij}^{\text{O}}\}, \{\hat{\alpha}_{ij}\}, \{\hat{\beta}_{ij}\}, \{x_{ij}^{\text{O}}\}) 
    &:= - \log P(\mathcal{X}|\Theta)
 \end{align}
 $\forall 1 \leq i \leq T$ and $1 \leq j \leq K $. 
 Where $\{\hat{x}_{ij}^{\text{O}}\}$ indicates the set of all pixel values over all the images in the dataset, similarly for others.
 In addition to the above uncertainty-aware fidelity loss term,
 the adversarial term depending upon $\mathcal{D}(\cdot; \theta^{\text{D}})$ is defined in terms of binary cross-entropy between the 
 true and predicted probability vectors for each of 
 the generated image (say $\hat{y}_i = \mathcal{D}(\hat{x}_i^\text{O}; \theta^\text{D})$) and 
 ground-truth image (say $y_i = \mathcal{D}(x_i^{\text{O}}; \theta^\text{D})$), respectively. The binary cross entropy loss is given by,
 $\mathcal{L}_{\text{CE}}(\hat{y}_i, y_i) = - \sum_{c} [ y_{ic} \log(\hat{y}_{ic}) + (1-y_{ic}) \log(1-\hat{y}_{ic}) ]$, where $y_{ic}$ represents the $c^{th}$ element in the output probability vector obtained from $\mathcal{D}$ with input $x^{\text{O}}_i$.
 Hence, $\mathcal{G}(\cdot; \theta^\text{G})$ minimizes the loss $\mathcal{L}_{\text{UG}}(\{\hat{x}_{ij}^{\text{O}}\}, \{\hat{\alpha}_{ij}\}, \{\hat{\beta}_{ij}\}, \{x_{ij}^{\text{O}}\})$ given by,
 \begin{align}
    \mathcal{L}_{\text{U}}(\{\hat{x}_{ij}^{\text{O}}\}, \{\hat{\alpha}_{ij}\}, \{\hat{\beta}_{ij}\}, \{x_{ij}^{\text{O}}\}) + 
    \lambda\sum_{i} \mathcal{L}_{\text{CE}}(\mathcal{D}(\hat{x}^{\text{O}}_i; \theta^{\text{D}}), 1)
 \label{eq:l_quest}
 \end{align}
 On the other hand, $\mathcal{D}(\cdot; \theta^{\text{D}})$ minimizes $\mathcal{L}_D(\{\hat{x}^{\text{O}}_i\}, \{x^{\text{O}}_i\})$ given by,
 \begin{align}
    \sum_i \mathcal{L}_{\text{CE}}( \mathcal{D}(x^{\text{O}}_i; \theta^{\text{D}}), 1) + \mathcal{L}_{\text{CE}}( \mathcal{D}(\hat{x}^{\text{O}}_i; \theta^{\text{D}}), 0)  
 \end{align}
 We use the strategy elucidated in \cite{nie2018medical,gan_tut} for training.
 
 The architecture of the proposed generator is inspired by U-Net \cite{ronneberger2015u}. We modify the U-Net such that the last convolutional layer is split into three, i.e., three independent convolutional layers attached to the penultimate block in U-Net (as shown in Figure~\ref{fig:model}). This model predicts $\hat{x}^O$, $\hat{\alpha}$, and $\hat{\beta}$.
 The discriminator network ( $\mathcal{D}(\cdot; \theta^{\text{D}})$ in Figure~\ref{fig:model} ) 
 is a typical CNN architecture as described in \cite{dar2019image}.
 Although image synthesis benefits from 3D processing, training CNNs with 3D images poses substantial challenges such as increased computational demand, cost, and training time. Several works use patch-based approaches to circumvent this issue. However, the final reconstruction
 from patches might result in artifacts \cite{mia_book}. In this work, we design our model to accept 2.5D input as described in \cite{subtlemed} and produce a 2.5D output, this allows the model to exploit 3D like neighborhOOD-noisy information while learning, resulting in better quality outputs. The final 3D output is obtained by joining (via averaging) overlapping 2.5D volumes as shown in Figure~\ref{fig:mergescheme}.
 
 \subsection{Training and Testing Scheme}
 \label{ssec:tt1}
 All the networks were trained using Adam optimizer \cite{adam} by sampling mini-batches of size 16. 
 The initial learning rate was set to $2e^{-4}$ and cosine annealing was used to decay the learning rate with epochs. 
 The $\lambda$ for MP and SR (Equation~\ref{eq:l_quest}) was set to $7e^{-4}$ and $1e^{-3}$ respectively.
 For numerical stability, the proposed network produces $\frac{1}{\hat{\alpha}}$ instead of $\hat{\alpha}$. 
 The positivity constraint on the output is enforced by applying the ReLU activation function at the end of the three output 
 layers in the network (Figure~\ref{fig:model}).
 In addition to the network-generated outputs ($\hat{x}^{\text{O}}$, $\hat{\alpha}$, $\hat{\beta}$), we also compute 
 aleatoric and epistemic uncertainty maps denoted by $\hat{\sigma}_{\text{aleatoric}}$ and $\hat{\sigma}_{\text{epistemic}}$, respectively.
 While $\hat{\sigma}_{\text{epistemic}}$ captures the uncertainty in the model parameters, $\hat{\sigma}_{\text{aleatoric}}$ captures 
 the data-dependent uncertainty, we combine both to produce $\hat{\sigma}$. 
 To capture the epistemic uncertainty, multiple forward passes ($R$ forward passes) are done for every single input image at inference with dropouts activated. 
 In this way, 
 the final output of the proposed framework, for a given input $x^{\text{I}}$ along with the uncertainty maps are given by,
 \begin{align}
    & \hat{x}^{\text{O}}_r = \mathcal{G}_{\text{O}}(\mathcal{G}_{\text{S}}(x^{\text{I}}; \theta^\text{S}_r);\theta^{\text{O}}_r)  \label{eq:in1} \\
    & \hat{\alpha}_r, \hat{\beta}_r =  \mathcal{G}_{\alpha}(\mathcal{G}_{\text{S}}(x^{\text{I}}; \theta^\text{S}_r);\theta^{\alpha}_r), 
    \mathcal{G}_{\beta}(\mathcal{G}_{\text{S}}(x^{\text{I}}; \theta^\text{S}_r);\theta^{\beta}_r) \label{eq:in2} \\
    & \hat{x}^{\text{O}}, \hat{\alpha}, \hat{\beta} = \frac{1}{R} \sum_{r=1}^{r=R} \hat{x}^{\text{O}}_r, \frac{1}{R} \sum_{r=1}^{r=R} \hat{\alpha}_r, 
      \frac{1}{R} \sum_{r=1}^{r=R} \hat{\beta}_r \label{eq:in3} \\
    & \hat{\sigma}_{\text{aleatoric}}^2, \hat{\sigma}_{\text{epistemic}}^2 = \frac{\hat{\alpha}^2 \Gamma(3/\hat{\beta})}{\Gamma(1/\hat{\beta})}, 
    \frac{\sum_{r=1}^{r=R} (\hat{x}^{\text{O}}_r - \hat{x}^{\text{O}})^2}{R} \label{eq:uncer_eq} \\
    & \hat{\sigma}^2 =  \hat{\sigma}_{\text{aleatoric}}^2 + \hat{\sigma}_{\text{epistemic}}^2 \label{eq:final_uncer}
 \end{align}
 where $r$ denotes the index for $r^{th}$ forward pass through the network 
 and $\{\theta^{\text{S}}_r, \theta^{\text{O}}_r, \theta^{\alpha}_r, \theta^{\beta}_r\}$ denotes the parameters that are instantiated due to dropouts on the $r^{th}$ forward pass, as described in \cite{gal}. In our experiments $R$ was set to 50.
 
 \section{Datasets and Experiments}
 \label{sec:rd}
 For the two applications QE and MP, we describe the {\em in vivo} training data and generation of OOD-noisy test data. 
 
 \subsection{Datasets}
 \label{sec:datasets}
 \begin{figure}
    \centering
    \includegraphics[width=0.47\textwidth]{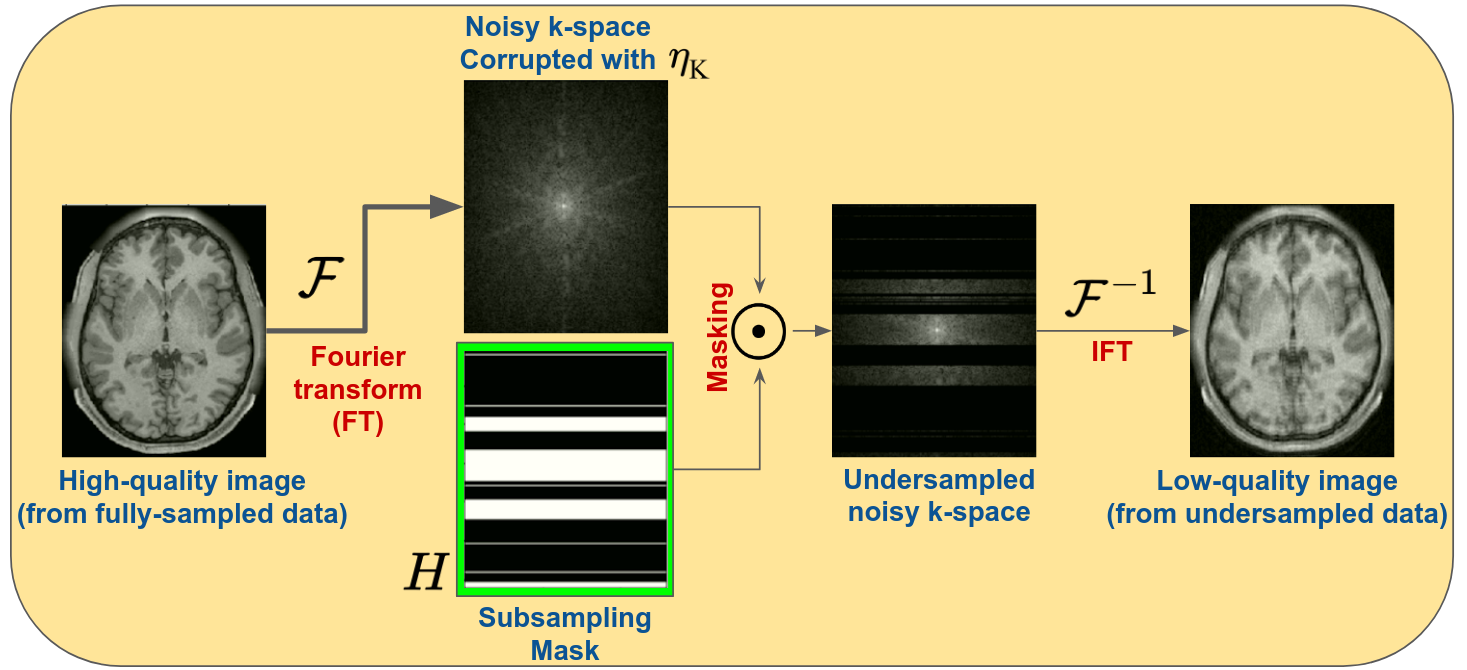}
    \caption
    {
    {\bf
    Retrospective data generation for QE.}. MRI undersampling is achieved by acquiring limited frequencies in k-space.
    }
    \label{fig:sr_schematic}
    \vspace{-15pt}
 \end{figure}
 We use the following two datasets: 
 (i)~proprietary dataset consisting of registered T1w MRI scans for QE, and
 (ii)~publicly available IXI dataset (\url{https://brain-development.org/ixi-dataset/}) for MP.
 The dataset in (i) consists of T1w MRI scans with $1$~mm $^3$ isotropic resolution from 100 subjects. We consider these images as the HQ-MR images for the QE experiment. 
 The methodology to obtain LQ-MR images is shown in Figure~\ref{fig:sr_schematic}. 
 For LQ data acquisition, we sample only $30\%$ 
 of the k-space using the mask in Figure~\ref{fig:sr_schematic}, resulting in an acceleration of $>3\times$.
 Subsequently, the LQ images are obtained as the inverse Fourier transform (IFT)
 of the undersampled k-space. 
 We use training, validation, and test split of 70, 10, and 20 non-overlapping subjects respectively.
 For each subject, we use 50 mid-brain axial slices.
 The dataset in (ii) consists of multi-contrast MR images from several healthy subjects, 
 collected across three different scanners. 
 For MP, we use T1w and T2w contrast images which were 
 co-registered (intra-subject) using ANTS \cite{ants}. 
 We employ a training, validation, and test split of 250, 50, 
 and 100 non-overlapping subjects respectively and
 we use 70 axial slices from the mid-brain region per subject.

 \subsection{Experiments}
 \label{ssec:exp}
 \begin{table}[!h]
    \vspace{-10pt}
    \centering
    \includegraphics[width=0.47\textwidth]{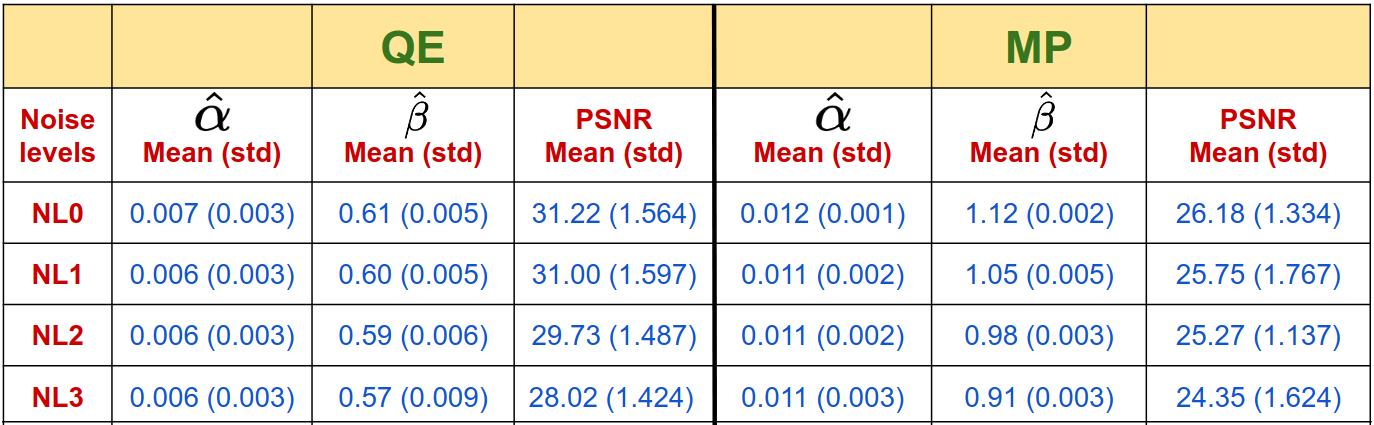}
    \caption
    {
    {\bf Trends across noise levels.} overall variations in scale parameter ($\hat{\alpha}$), beta parameter ($\hat{\beta}$), and the output PSNR values across all the noise levels for QE and MP. 
    }
    \label{fig:ab_table}
    \vspace{-4pt}
 \end{table}
 
 \begin{figure*}
    \centering
    \includegraphics[width=\textwidth]{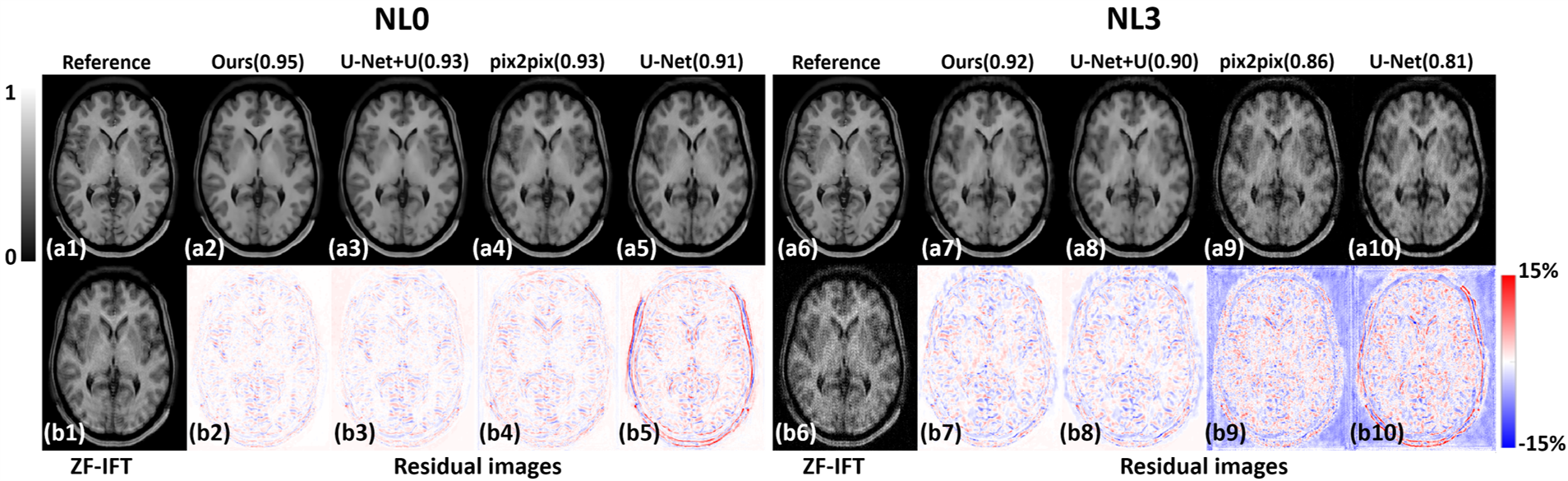}
    
    \caption
    {
    {\bf 
    Results for QE.
    }
    Predicted HQ images with input LQ images at
    two different noise-levels (NLs),
    \textbf{(b1):}~NL0 (high SNR, same NL as training-set),
    and \textbf{(b5):}~NL3 (low SNR, highest simulated NL). 
    {\bf (a2)--(a5) and (a7)--(a10):}
    Predicted images for all the methods at NL0 and NL3, respectively.
    {\bf (b2)--(b5) and (b7)--(b10):} Corresponding residual images.
    SSIM values are indicated within paranthesis.
    }
    \label{fig:vis_ass_sr}
    \vspace{-8pt}
 \end{figure*}
 \begin{figure*}
    \includegraphics[width=\textwidth]{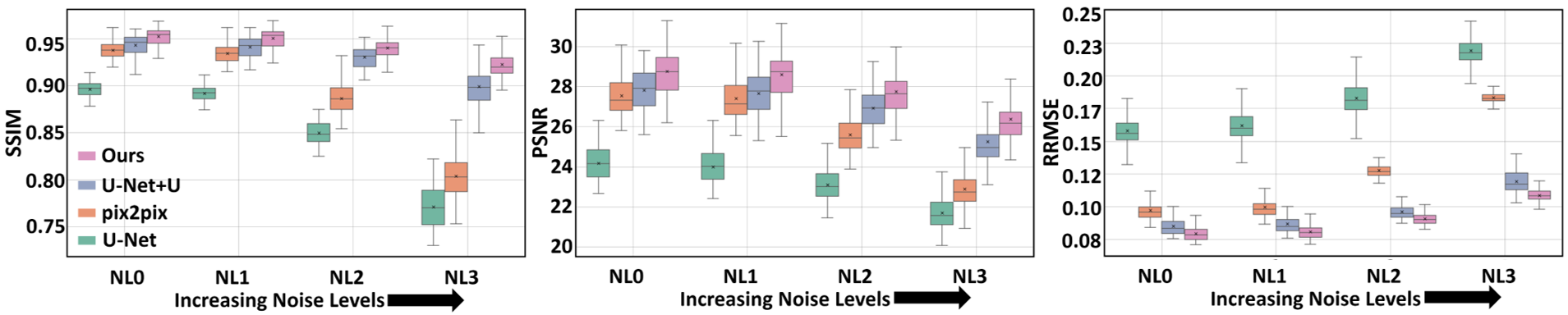}
    \caption
    {
    {\bf Quantitative assessment for QE.} SSIM, PSNR, and RRMSE values for all the methods at multiple noise-levels (NL0 to NL3).
    At each NL, 50 mid-brain slices from each subject (20 test-subjects) were evaluated (i.e., 1000 slices).
    }
    \label{fig:quantitative_assessment_sr}
    \vspace{-8pt}
 \end{figure*}
 
 \textbf{OOD-noisy data in MRI.} 
 MRI data acquisition depends on several factors that may affect the quality of the MRI scans leading to a wide distribution of datasets. In practice, the dataset used for learning algorithms will not cater to all the possible variations in the distributions; hence, it is imperative to design learning algorithms that are robust to a wide-range of OOD-noisy data.
 In particular, the signal-to-noise ratio (SNR) of the MRI k-space data (and consequently the image) depends on various factors such as field strength, pulse sequence, tissue characteristics, number of receiver coils and their sensitivities, scan/physiological parameters, etc.
 ~\cite{empEffectGauss_lustwig,zhu2020removal,robson2008comprehensive,lustig2007sparse}. 
 Hence, for both QE and MP, we generate OOD-noisy MRI data that captures variations in noise levels as described below.
 We consider the noise-level (NL) of the training set to be NL0 for both QE and MP. We model \textit{three} increasing noise-levels, NL1--NL3, of OOD-noisy degradations for the test data. Degradations are in (i)~k-space for QE and (ii)~image-space for MP.
 \newline\indent
 \textbf{QE.} 
 Having trained the network using images at NL0, at inference, we evaluate the robustness of all the methods at {\em three} increasing levels of noise in the k-space. 
 In practice, this is obtained by increasing the magnitude of complex Gaussian noise $\eta_{\text{K}}$ in the k-space and as described in Equation~\ref{eq:k_corrupt}. In our experiments, sampling mask $H$ is fixed (i.e., fixing the k-space trajectory~\cite{mezrich1995perspective,circus}, Figure~\ref{fig:sr_schematic}), acquiring only $30\%$ of the k-space.
 The PSNR values of the LQ images at NL0--NL3, averaged across the test set, were around $\{21, 18, 16, 14\}$ dB with respect to HQ images.
 \newline\indent
 \textbf{MP.} 
 Here we employ two kinds of OOD-noisy test data: 
 (i)~evaluation at {\em three} increasing levels of noise in the image-space and 
 (ii)~inclusion of \textit{unseen} synthetic lesions. 
 The image-space degradations 
 are obtained by increasing the magnitude of Gaussian noise $\eta_{\text{I}}$ 
 Equation~\ref{eq:mri_corrupt}. 
 For the lesions related experiment,
 we add lesions in both the input (T1w) and the reference images (T2w) based on the segmented lesion masks available from the BRATS 2020 dataset (\url{https://www.med.upenn.edu/cbica/brats2020/data.html}).
 \vspace{-16pt}
 \section{Results and Discussion}
 \vspace{-5pt}
 In this work, for a fair evaluation, 
 we modify the baselines (originally in 2D) to use a 2.5D-style training strategy.
 We use structural similarity index (SSIM)~\cite{ssim}, peak signal-to-noise ratio (PSNR) and relative root mean squared error (RRMSE) for quantitative comparison. 
 RRMSE between two images $a$ and $b$ is defined as $\text{RRMSE}(a,b) = \|a-b\|_F/\|a\|_F$, where $\|\cdot\|_F$ indicates Frobenius norm.
 
 {\bf Quantitative and Qualitative Evaluations for QE.} 
 We use the following state-of-the art methods for comparison: 
 (i) \textbf{U-Net} as used in~\cite{fastMRI},
 (ii) \textbf{pix2pix} as used in~\cite{fastMRI,pix2pix}, and
 (iii) \textbf{U-Net with uncertainty (U-Net+U)} as used in~\cite{fastMRI}. 
 Finally, our proposed method, \textbf{URGAN (ours)}.
 Our uncertainty model is based on \textit{generalized Gaussian distributions} allowing our loss function to be tuned automatically, unlike that of (U-Net+U) as described in~\cite{fastMRI}. 
 All the baseline models are trained with the loss function proposed in their respective formulations.
 \begin{figure*}
    \centering
    \includegraphics[width=0.88\textwidth]{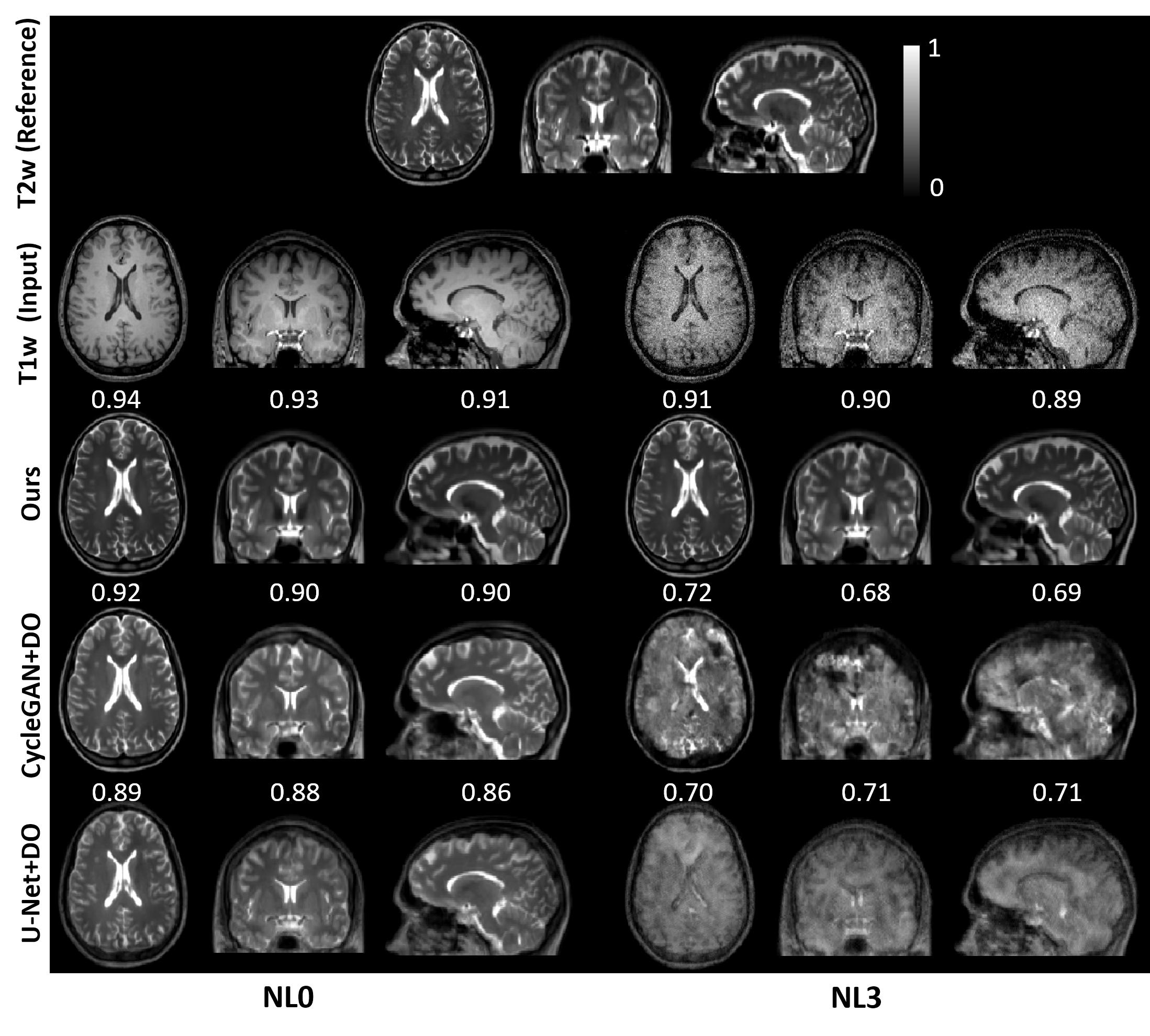}
    \vspace{-6pt}
    \caption
    {
    {\bf 
    Qualitative assessment for MP using orthogonal views.
    } 
    Results on two different input noise-levels (NLs): NL0 (High SNR, same NL as training-set), NL3 (Low SNR, highest simulated NL). SSIM values for each slice embedded.
    }
    \label{fig:ortho_mp}
    \vspace{-13pt}
 \end{figure*}
 Figure \ref{fig:vis_ass_sr} shows the predicted images for QE at NL0 and NL3, for a representative test slice.
 At NL0,  
 the predicted images from all the methods (Figure~\ref{fig:vis_ass_sr} (a2) -- (a5)) have low residual error and are comparable among each other.  However, at NL3, pix2pix and U-Net generate images with artifacts (Figure~\ref{fig:vis_ass_sr} (b4) - (b5)). 
 Whereas, the predicted images from methods that model uncertainty (Ours and U-Net+U) show a superior recovery of structure and contrast in addition to the removal of artifacts (Figure~\ref{fig:vis_ass_sr} (a2-a3 and a7-a8)).
 The quantitative evaluation across all the OOD-noisy test data (Figure~\ref{fig:quantitative_assessment_sr}) shows that the performance of all the baseline methods suffers substantially at higher degradation levels, emphasizing the improved robustness due to the proposed GGD-based uncertainty model.
 Table~\ref{fig:ab_table} shows that with increasing NLs, the mean of $\hat{\beta}_{ij}$s decreases, explained in~\ref{sec:uncer_sec}.
 %
 \newline
 \indent
 {\bf Qualitative and Qualitative Evaluations for MP.}
 We use the following state-of-the-art methods for comparison:
 (i)~\textbf{U-Net} as used in~\cite{chartsias2017multimodal,subtlemed} that is adapted to work with unimodal input (T1w) and predicts the unimodal output (T2w).
 (ii)~\textbf{CycleGAN} from \cite{dar2019image}. Here we used a U-Net for the generator (as it led to better performance). 
 (iii)~We improve the robustness of U-Net (from \cite{chartsias2017multimodal,subtlemed}) to OOD-noisy noisy input data by employing dropouts (DO) during both training and inference, called \textbf{U-Net+DO}.
 (iv)~Similarly, we employ dropouts for CycleGAN (from \cite{dar2019image}) giving us \textbf{CycleGAN+DO}.
 We perform multiple forward passes and take the mean to get the final outputs of models with dropouts activated at inference.
 Both U-Net and U-Net+DO are trained using the pixel-wise $\ell_1$ loss; CycleGAN and CycleGAN+DO use 
 an additional adversarial loss with cycle consistency as in \cite{dar2019image}.
 \begin{figure*}
    \centering
    \includegraphics[width=\textwidth]{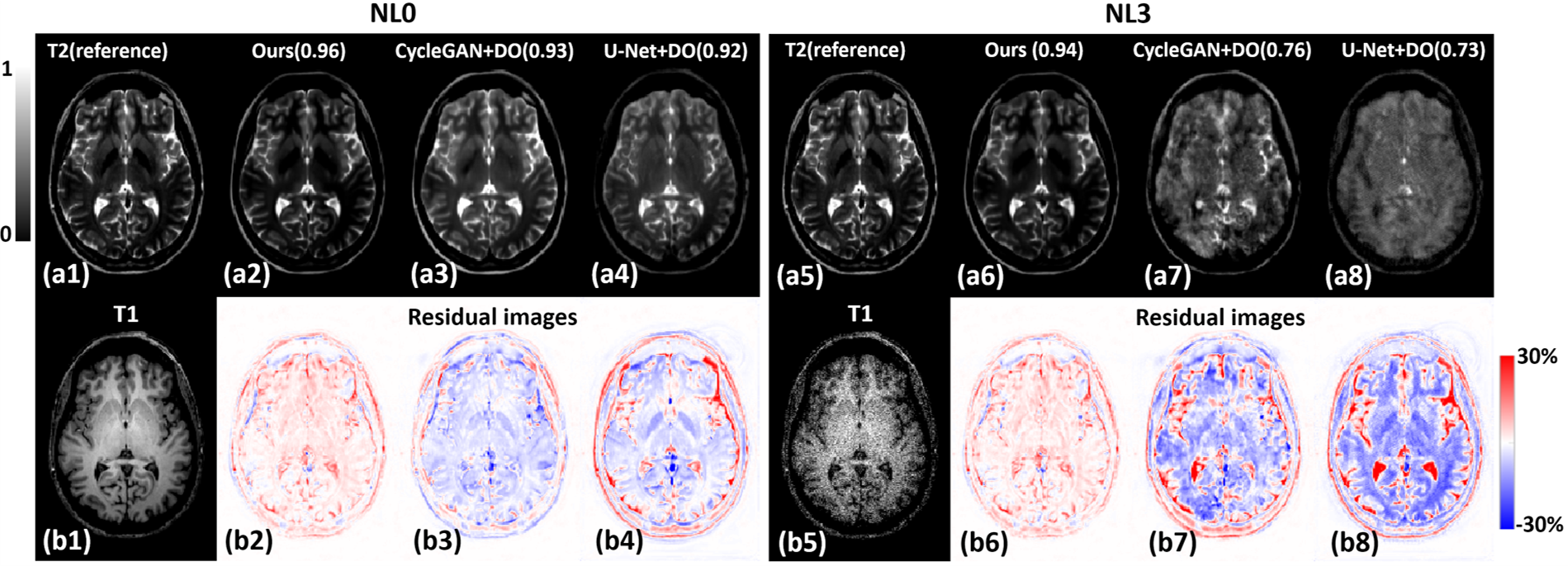}
    \caption
    {
    {\bf 
    Results for MP (axial view).
    }
    Predicted T2w images with input T1w images at
    two different noise-levels (NLs),
    \textbf{(b1):}~NL0 (high SNR, same NL as training-set),
    and \textbf{(b5):}~NL3 (low SNR, highest simulated NL). 
    {\bf (a2)--(a4) and (a6)--(a8):}
    Predicted (T2w) for all the methods at NL0 and NL3, respectively.
    {\bf (b2)--(b4) and (b6)--(b8):} Corresponding residuals.
    SSIM values are indicated within paranthesis.
    }
    \label{fig:vis_ass_mp}
    \vspace{-5pt}
 \end{figure*}
 \begin{figure*}
    \includegraphics[width=\textwidth]{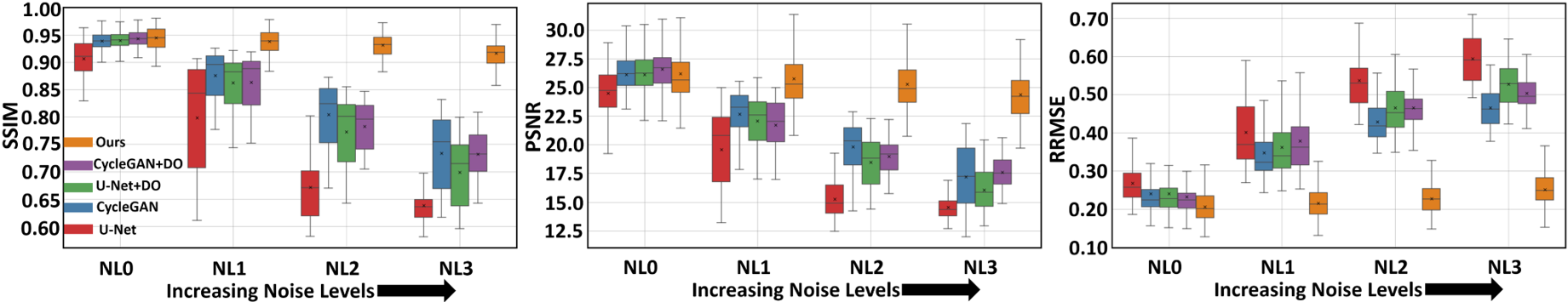}
    \caption
    {
    {\bf Quantitative assessment for MP.} SSIM, PSNR, and RRMSE values for all the methods at multiple noise-levels (NL0 to NL3).
    At each NL, 70 mid-brain slices from each subject (100 test-subjects) were evaluated (i.e., 7000 slices).
    }
    \label{fig:quantitative_assessment_mp}
    \vspace{-10pt}
 \end{figure*}
 \newline
 \indent
 Figure~\ref{fig:ortho_mp} shows the MP orthogonal views through the 3D output obtained from our method as well as the best performing baselines. U-Net+DO and CycleGAN+DO show improved performance in comparison to the non-DO counterparts as evident from Figure~\ref{fig:quantitative_assessment_mp}. 
 Similar to QE,
 at NL0, outputs of all methods are comparable. However, at higher NLs, the proposed method outperforms all the baselines substantially, across all metrics (Figure~\ref{fig:ortho_mp} and \ref{fig:quantitative_assessment_mp}).
 Figure~\ref{fig:vis_ass_mp} shows the predicted images for a representative test axial input slice along with the residuals for the proposed model, U-Net+DO, and CycleGAN+DO at two different NLs (NL0 and NL3).
 At NL0, 
 the predicted images from all the methods (Figure~\ref{fig:vis_ass_mp} (a2) -- (a4) appear closer in structure and contrast to the ground truth (Figure~\ref{fig:vis_ass_mp} (a1)).  Our method shows the least residual (Figure~\ref{fig:vis_ass_mp} (b2)) compared to that of CycleGAN+DO (b3) and U-Net+DO (b4). 
 At NL3,
 our method (Figure~\ref{fig:vis_ass_mp} (a6)) outperforms both CycleGAN+DO (Figure~\ref{fig:vis_ass_mp} (a7)) and U-Net+DO (Figure~\ref{fig:vis_ass_mp} (a8)) in terms of visual quality as well as SSIM values. All baselines show poor synthesis of structure, contrast, and limited removal of noise. Our method removes significant amount of noise, recover structure, contrast and texture, and is closer in appearance to the reference image (Figure~\ref{fig:vis_ass_mp} (a5)). 
 
 \begin{figure*}[!h]
    \centering
        \includegraphics[width=0.93\textwidth]{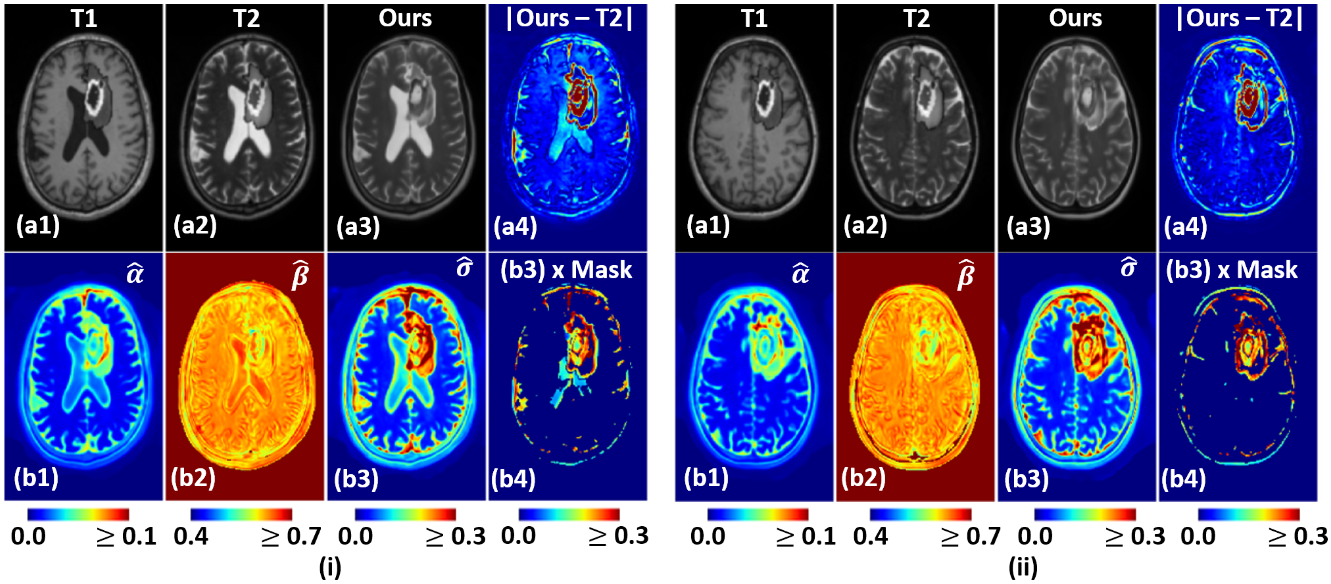}
    
    \caption
    {
    {\bf Uncertainty quantification on OOD-noisy data (synthetic lesion added to MP test data).}
     {\bf Subfigures (i) -- (ii)} show results from two representative slices.
     {\bf (a1)-(a2):} Input T1w ($x^{\text{T1}}$) and the corresponding reference T2w ($x^{\text{T2}}$).
     {\bf (a3)-(a4):} Predicted image ($\hat{x}^{\text{T2}}$) and absolute error map.
     {\bf (b1)-(b2):} The learned scaling ($\hat{\alpha}$) and shape ($\hat{\beta}$) parameter maps.
     {\bf (b3):} Uncertainty map ($\hat{\sigma}$).
     {\bf (b4):} Masked aleatoric uncertainty map (refer Section~\ref{sec:uncer_sec}). 
    }
    \label{fig:lesion_mp}
    \vspace{-8pt}
 \end{figure*}
 \begin{figure*}
    \centering
    \includegraphics[width=0.94\columnwidth]{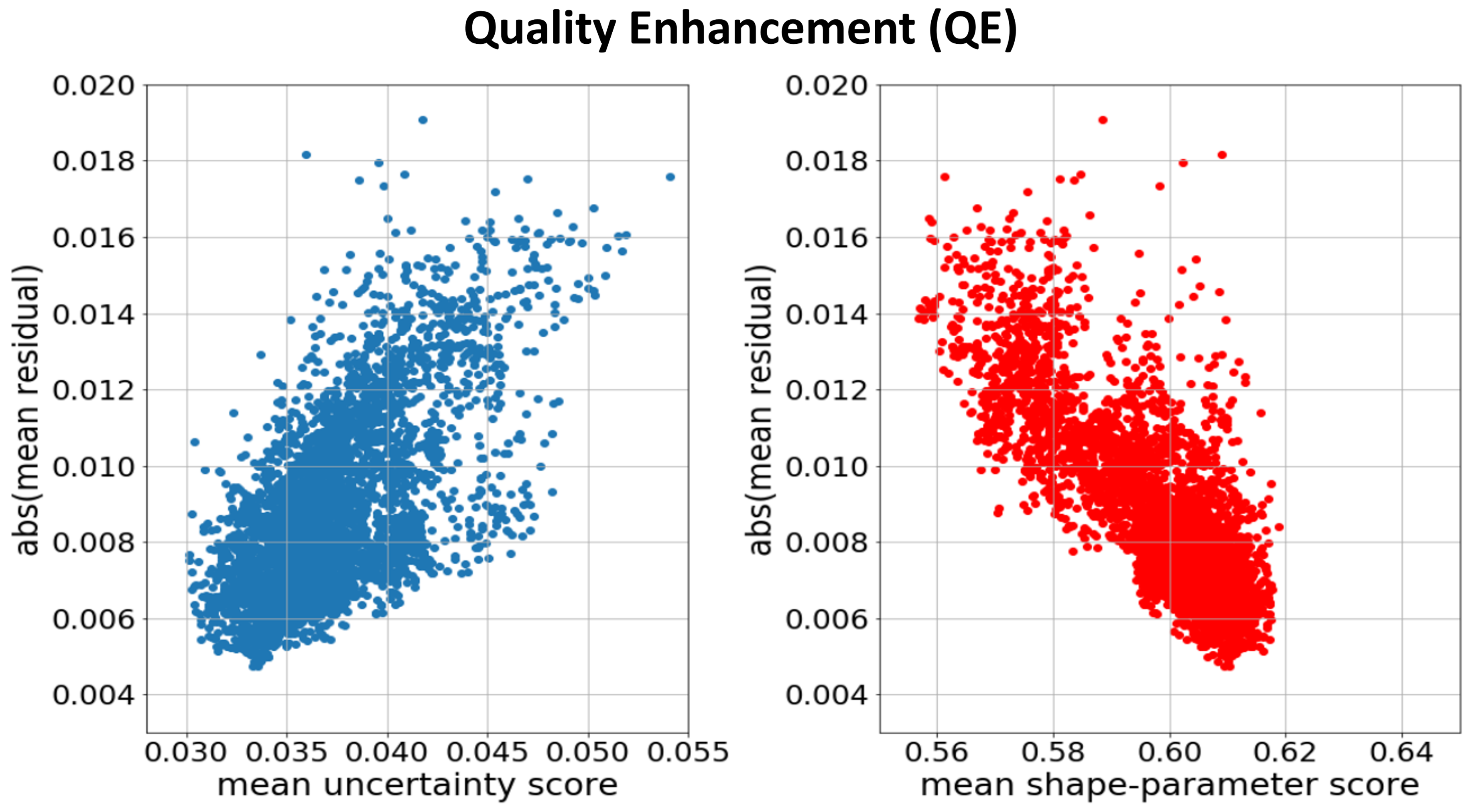}
    \includegraphics[width=0.94\columnwidth]{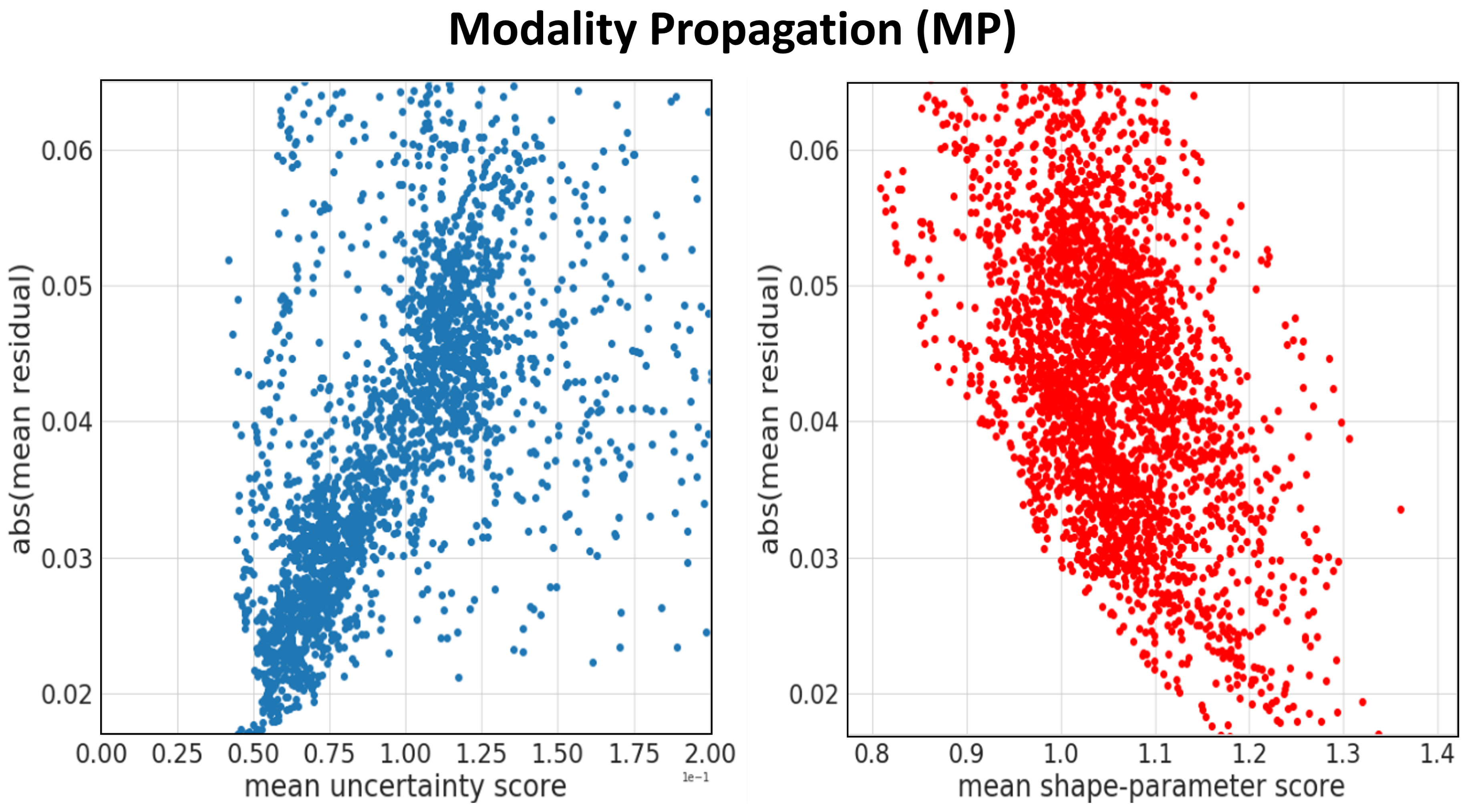}
    \caption{{\bf Correlation between residual, uncertainty, and shape-parameter. Data from all NLs.}
    Each point indicates an image, uncertainty score and shape-parameter score is calculated for an image by taking mean of 
    $\hat{\sigma}$ and $\hat{\beta}$ over all the pixels.
    }
    \vspace{-15pt}
    \label{fig:corr}
 \end{figure*}
 
 \subsection{Uncertainty quantification}
 \label{sec:uncer_sec}
 Along with improved accuracy of the predicted images, we demonstrate the efficacy of estimating uncertainty maps, ($\hat{\sigma}$, $\hat{\sigma}_{\text{aleatoric}}$, $\hat{\sigma}_{\text{epistemic}}$).
 Uncertainty maps are derived from the outputs of the network: $\hat{x}^{\text{O}}$, $\hat{\alpha}$, and $\hat{\beta}$ as described in Equation~\ref{eq:uncer_eq}-\ref{eq:final_uncer}.
 Scatter plots in Figure~\ref{fig:corr} show that for both QE and MP, the residuals between the predictions and the ground-truths correlate positively with the uncertainty obtained from our framework, i.e., high overall residuals correspond to high overall uncertainty for the image. 
 Therefore, higher overall uncertainty may be used as a proxy to infer about increased imperfections in reconstruction/synthesis.
 Interestingly, Figure~\ref{fig:corr} also show that high residuals correlate negatively with predicted shape-parameter ($\hat{\beta}$) that is 
 consistent with the fact that lower shape-parameter values correspond to heavy-tailed distributions for residuals, i.e., outliers.
 \newline
 \indent We also employ a testing scenario for MP, where the incoming test subject has an unseen pathology (a lesion).
 In this case too, we use the network which was trained on healthy individuals only.
 Figures~\ref{fig:lesion_mp} (a1) and (a2) show the input (T1w) and the reference (T2w) image with simulated lesions. 
 The sub-figures (i) and (ii) show results on two representative slices.
 The predicted image ($\hat{x}^{\text{T2}}$, Figure~\ref{fig:lesion_mp} (a3)), the scale parameter of the GGD 
 ($\hat{\alpha}$, Figure~\ref{fig:lesion_mp} (b1)), and the corresponding shape parameter ($\hat{\beta}$, Figure~\ref{fig:lesion_mp} (b2)), 
 are the outputs of the network.
 The derived uncertainty map is shown in Figure~\ref{fig:lesion_mp} (b3), obtained using Equations~\ref{eq:in3} -- \ref{eq:final_uncer}. 
 For the background region, both the scale and shape maps show values that are less spatially varying.
 While the scale map ($\hat{\alpha}$) captures edges in the image at a coarser level, 
 the shape map ($\hat{\beta}$) captures subtle features in the image. In all the slices intensity values in the $\hat{\beta}$ image, within the lesion region are significantly lower than other regions, indicating outliers.  
 We observe that the absolute residual image has peak values in and around the lesion. This is expected as the training data is devoid of such pathological cases. 
 The regions with ``high'' intensity values within the uncertainty map ($\hat{\sigma}$, Figure~\ref{fig:lesion_mp} (b3)) show high residual values as well (Figure~\ref{fig:lesion_mp} (a4)).
 In addition to showing peak values of the uncertainty maps in regions with high residual error, the uncertainty map is able to capture 
 minor variations in the predicted images, especially at the skull boundary and gyri. 
 To make the uncertainty map clinically relevant, we identify ``high'' residual values as the ones that are above a pre-defined threshold, say $\tau$. The value of $\tau$ is fixed as follows: we choose $\tau$ to be slightly above the standard deviation of the highest noise-level at which the network provided reasonably accurate reconstructed images (NL3).
 In this paper, we choose $\tau = 0.17$, which is slightly above the estimated standard deviation of the noise at NL2. 
 Figure~\ref{fig:lesion_mp} (b4) shows the product of the uncertainty map ($\hat{\sigma}$) and the binary mask obtained by thresholding the absolute error map at $\tau = 0.17$.
 Clearly, the resultant uncertainty map (Figure~\ref{fig:lesion_mp} (b4)) reflects pixels with the highest predicted uncertainty. 
 Hence, uncertainty map can be a proxy for residual map
 (\textit{that is not available at inference}).
 \newline
 \indent\textbf{Conclusion.}
 In this work, we proposed a GAN-based framework with adaptive quasi-norm loss functions for improved robustness to unseen perturbations on test data. We demonstrated the efficacy of our network on two key applications arising in the field of medical imaging, namely undersampled MRI reconstruction and modality propagation.
 Enhanced output produced using this method after post-processing can serve in clinical decision making based.
 We compared our network with state-of-the-art networks for both applications and found that the performance of our network is substantially better (both qualitatively and quantitatively) at high noise levels, and comparable at noise levels similar to the training set. While our experiments with perturbations related to scanners (noise level study) showed that our framework can generate predicted images with low residual, 
 the experiments related to physiological perturbations (lesion study) showed that our uncertainty maps can be proxy for the residual maps.
 
 {\small
 \bibliographystyle{ieee_fullname}
 \bibliography{paper}
 }
 
 \end{document}